\documentclass[letter]{aa}
\usepackage[varg]{txfonts}
\usepackage{graphicx}

\begin{document}
   \title{Four temporary Neptune co-orbitals: (148975) 2001 XA$_{255}$, (310071) 
          2010 KR$_{59}$, (316179) 2010 EN$_{65}$ and 2012 GX$_{17}$
          \thanks{Tables \ref{L4T1}-\ref{L5T} and Figures \ref{L4a}-\ref{L5} are 
                  only available in electronic form via http://www.edpsciences.org}
         } 
   \author{C. de la Fuente Marcos
           \and
           R. de la Fuente Marcos}
   \authorrunning{C. de la Fuente Marcos \and R. de la Fuente Marcos}
   \titlerunning{Neptune co-orbitals: 2001 XA$_{255}$, 2010 KR$_{59}$, 
                 2010 EN$_{65}$ and 2012 GX$_{17}$}
   \offprints{C. de la Fuente Marcos, \email{nbplanet@fis.ucm.es}
              }
   \institute{Universidad Complutense de Madrid,
              Ciudad Universitaria, E-28040 Madrid, Spain}
   \date{Received 13 September 2012 / Accepted 10 October 2012}

   \abstract
      {Numerical simulations suggest that Neptune primordial co-orbitals may 
       significantly outnumber the equivalent population hosted by Jupiter, yet 
       the objects remain elusive. Since the first discovery in 2001 just ten 
       minor planets, including nine Trojans and one quasi-satellite, have been 
       positively identified as Neptune co-orbitals. In contrast, Minor Planet
       Center (MPC) data indicate that more than 5,000 objects are confirmed 
       Jupiter co-orbitals. On the other hand, some simulations predict that a 
       negligible fraction of passing bodies are captured into the 1:1 
       commensurability with Neptune today. 
       }
      {Hundreds of objects have been discovered in the outer solar system during 
       the various wide-field surveys carried out during the past decade, and
       many of them have been classified using cuts in the pericentre and other
       orbital elements. This leads to possible misclassifications of resonant 
       objects. Here, we explore this possibility to uncover neglected Neptune 
       co-orbitals. 
       }  
      {Using numerical analysis techniques, we singled out eleven candidates and 
       used $N$-body calculations to either confirm or reject their co-orbital 
       nature. 
       }
      {We confirm that four objects previously classified as Centaurs by the 
       MPC currently are temporary Neptune co-orbitals. (148975) 2001 XA$_{255}$ 
       is the most dynamically unstable of the four. It appears to be a 
       relatively recent (50 kyr) visitor from the scattered disk on its way to 
       the inner solar system. (310071) 2010 KR$_{59}$ is following a 
       complex horseshoe orbit, (316179) 2010 EN$_{65}$ is in the process of
       switching from L$_4$ to L$_5$ Trojan, and 2012 GX$_{17}$ is a promising 
       L$_{5}$ Trojan candidate in urgent need of follow-up. The four objects 
       move in highly inclined orbits and have high eccentricities. These 
       dynamically hot objects are not primordial 1:1 librators, but are 
       captured and likely originated from beyond Neptune, having entered the 
       region of the giant planets relatively recently.
       }
      {Casting doubt over claims by other authors, our results show that Neptune 
       can still efficiently capture co-orbitals for short periods of time and 
       that the cuts in the orbital elements are unreliable criteria to classify 
       objects orbiting in the outer solar system. As in the case of Jupiter 
       Trojans, our results suggests that Neptune's L$_5$ point is less stable 
       than L$_4$, in this case perhaps due to the influence of Pluto.
       }


   \maketitle

   \section{Introduction}
      Numerical simulations predict that Neptune may have retained a significant amount of its primordial co-orbital minor planet
      population, including Trojans and quasi-satellites (Holman \& Wisdom 1993; Wiegert et al. 2000; Nesvorn\'y \& Dones 2002; 
      Marzari et al. 2003; Kortenkamp et al. 2004; Dvorak et al. 2007; Nesvorn\'y \& Vokrouhlick\'y 2009; Zhou et al. 2009, 2011; 
      Lykawka et al. 2009, 2010, 2011). The first Neptune Trojan, 2001 QR$_{322}$, was serendipitously discovered by Chiang et al. 
      (2003). 2004 UP$_{10}$, 2005 TN$_{53}$ and 2005 TO$_{74}$ followed (Sheppard \& Trujillo 2006). Then 2006 RJ$_{103}$ and 
      2007 VL$_{305}$, all of them leading 60$^{\circ}$ ahead of Neptune (i.e. they are L$_4$ Trojans). Shortly after, the first 
      L$_5$ Trojan, 2008 LC$_{18}$, was found (Sheppard \& Trujillo 2010), followed by the second one, 2004 $KV_{18}$. These eight 
      Trojans have recently been joined by one short-term quasi-satellite, (309239) 2007 RW$_{10}$ (de la Fuente Marcos \& de la 
      Fuente Marcos 2012a) and yet another L$_5$ Trojan, 2011 HM$_{102}$, (Tholen et al. 2012; Parker et al. 2012). Confirmed 
      Jupiter co-orbitals currently amount to more than 5,000 objects\footnote{http://www.minorplanetcenter.net/iau/lists/JupiterTrojans.html}; 
      theoretical and numerical expectations predict even larger numbers for Neptune, although they are harder to detect because 
      they are farther away. On the other hand, it is commonly thought that Neptune cannot efficiently capture objects into the 
      1:1 commensurability even for short periods of time (Horner \& Evans 2006). 

      During the past decade, wide-field surveys of the outer solar system have found hundreds of objects passing in the 
      neighbourhood of Neptune, many of which were classified using cuts in the pericentre, $q$, and other orbital elements 
      (for example, the MPC defines that Centaurs must have a perihelion larger than Jupiter's orbit and a semi-major axis shorter 
      than Neptune's). This leads to misclassifications of resonant objects (Shankman 2012). It is possible that some of these 
      objects may have not been properly identified and that they are, in fact, trapped (even if temporarily) in a 1:1 
      commensurability with Neptune. Here, we explore this possibility and try to uncover misidentified Neptune co-orbitals. In 
      this letter, we use $N$-body simulations to confirm the possible co-orbital nature with Neptune of a number of objects 
      currently classified as Centaurs by the Minor Planet Center (MPC). The numerical model is described in the next section and 
      the results are presented in Sections 3 to 6. The results are discussed and our conclusions summarized in Section 7.

   \section{Candidates and numerical experiments}
      Following Mikkola et al. (2006) and to study the librational properties of any possible candidates, we define the relative 
      deviation of the semi-major axis from that of Neptune by $\alpha = (a - a_N) / a_N$, where $a$ and $a_N$ are the semi-major 
      axes of the object and Neptune, respectively, and also the relative mean longitude $\lambda_r = \lambda - \lambda_N$, where 
      $\lambda$ and $\lambda_N$ are the mean longitudes of the object and Neptune, respectively. If $\lambda_r$ oscillates around 
      0$^{\circ}$, we call the object a quasi-satellite; a Trojan is characterized by $\lambda_r$ oscillating around +60$^{\circ}$ 
      (L$_4$ Trojan) or -60$^{\circ}$ (or 300$^{\circ}$, L$_5$ Trojan); finally, an object oscillating with amplitude 
      $> 180^{\circ}$ follows a horseshoe orbit in a frame of reference rotating with Neptune (see, e.g., Murray \& Dermott 1999). 
      Objects that switch libration from the Lagrangian point L$_4$ to L$_5$ (or vice versa) are called jumping Trojans (Tsiganis 
      et al. 2000). Searching for Neptune co-orbital candidates among known minor bodies with relative semi-major axis $|a - a_{N}| 
      <$ 2 AU, we found 21 objects, 10 of which were already documented as co-orbitals. We regarded the remaining 11 as candidates 
      and then used $N$-body calculations to either confirm or reject their putative co-orbital nature. The orbit computations 
      were completed using a Hermite integration scheme (Makino 1991; Aarseth 2003). The $N$-body code is publicly available from 
      the IoA web site\footnote{http://www.ast.cam.ac.uk/$\sim$sverre/web/pages/nbody.htm}. Our calculations include the 
      perturbations by the eight major planets, treating the Earth and the Moon as two separate point masses, the barycentre of 
      the Pluto-Charon system, and the three largest asteroids. For accurate initial positions and velocities we used the 
      heliocentric ecliptic Keplerian elements provided by the JPL on-line solar system data service\footnote{http://ssd.jpl.nasa.gov/?planet\_pos} 
      (Giorgini et al. 1996) and initial positions and velocities based on the DE405 planetary orbital ephemerides (Standish 1998) 
      referred to the barycentre of the solar system. In addition to the orbital calculations completed using the nominal elements 
      in Table \ref{elements}, we have performed 50 control simulations (for each object) with sets of orbital elements obtained 
      from the nominal ones and the quoted uncertainties (3$\sigma$) when available. Additional details can be found in de la 
      Fuente Marcos \& de la Fuente Marcos (2012b). We have also validated our simulations against previous work by integrating 
      the orbits of the known nine Neptune Trojans (orbital elements in Tables \ref{L4T1}-\ref{L5T}), see Figs. \ref{L4a}-\ref{L5}; 
      results are consistent with those from other authors.
%
%
         \begin{table*}
          \fontsize{8}{10pt}\selectfont
          \tabcolsep 0.35truecm
          \caption{Heliocentric Keplerian orbital elements of the objects studied in this research.
                   Values include the 1-$\sigma$ uncertainty when available.
                   (Epoch = JD2456200.5, 2012-Sep-30.0 except for 2012 GX$_{17}$ that is JD2456020.5, 2012-Apr-3.0; J2000.0 ecliptic 
                   and equinox. Sources: JPL Small-Body Database and AstDyS-2.)
                  }
          \begin{tabular}{cccccc}
           \hline
                                                        &   & (148975) 2001 XA$_{255}$ & (310071) 2010 KR$_{59}$ & (316179) 2010 EN$_{65}$ & 2012 GX$_{17}$ \\
           \hline
            semi-major axis, $a$ (AU)                              & = & 28.8587$\pm$0.0010    & 29.970$\pm$0.004     & 30.719$\pm$0.002     & 30.1330172 \\
            eccentricity, $e$                                      & = & 0.676579$\pm$0.000010 & 0.56676$\pm$0.00006  & 0.31492$\pm$0.00004  & 0.4149134  \\
            inclination, $i$ ($^{\circ}$)                          & = & 12.62072$\pm$0.00002  & 19.67387$\pm$0.00005 & 19.24716$\pm$0.00003 & 35.29849   \\
            longitude of the ascending node, $\Omega$ ($^{\circ}$) & = & 105.92560$\pm$0.00014 & 46.7483$\pm$0.0003   & 234.2657$\pm$0.0003  & 207.75190  \\
            argument of perihelion, $\omega$ ($^{\circ}$)          & = & 90.3433$\pm$0.0004    & 108.2697$\pm$0.0010  & 225.236$\pm$0.003    & 94.00361   \\
            mean anomaly, $M$ ($^{\circ}$)                         & = & 5.3117$\pm$0.0002     & 5.6713$\pm$0.0014    & 36.831$\pm$0.005     & 298.44011  \\
            perihelion distance, $q$ (AU)                          & = & 9.33350$\pm$0.00010   & 12.9842$\pm$0.0008   & 21.0450$\pm$0.0009   & 17.6304246 \\
            absolute magnitude, $H$ (mag)                          & = & 11.2                  & 7.7                  & 6.9                  & 7.8        \\
           \hline
          \end{tabular}
          \label{elements}
         \end{table*}
%
%
%
%
     \begin{figure}
       \centering
        \includegraphics[width=\linewidth]{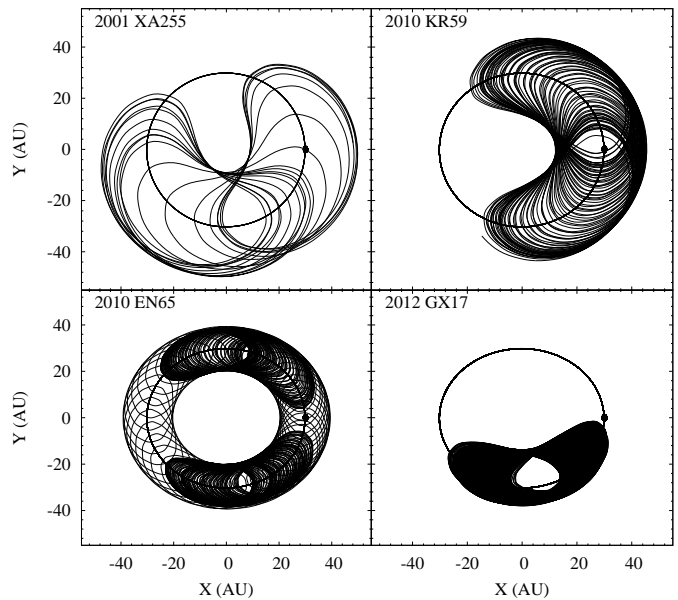}
        \caption{Motion of the objects in a coordinate system rotating with Neptune. The orbit and position of Neptune 
                 are also plotted. The time intervals displayed are 2001 XA$_{255}$ (-1.5, 1.5) kyr, 2010 KR$_{59}$ (5, 20) kyr, 
                 2010 EN$_{65}$ (-20, 20) kyr, and 2012 GX$_{17}$ (-50, 50) kyr. 
                }
        \label{4orbits}
     \end{figure}
%
%
%
%
     \begin{figure}
       \centering
        \includegraphics[width=\linewidth]{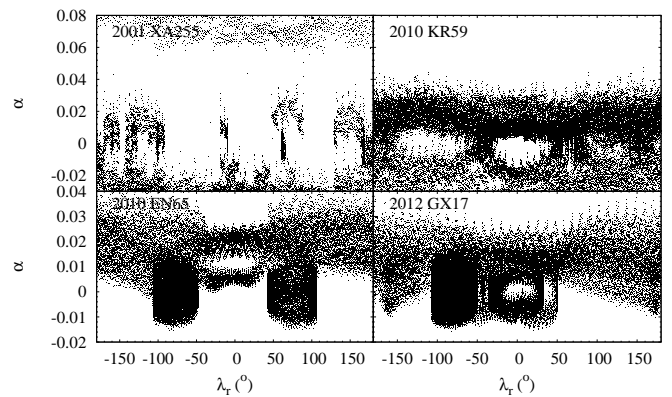}
        \caption{Resonant evolution. The $\alpha$ parameter (see text for details) as a function of the relative mean 
                 longitude during the time interval (-50, 50) kyr, except for 2012 GX$_{17}$, (-100, 100) kyr.
                }
        \label{4alpha}
     \end{figure}
%
%

   \section{(148975) 2001 XA$_{255}$: very dynamically unstable}
\begin{figure*}
       \centering
        \includegraphics[width=\textwidth]{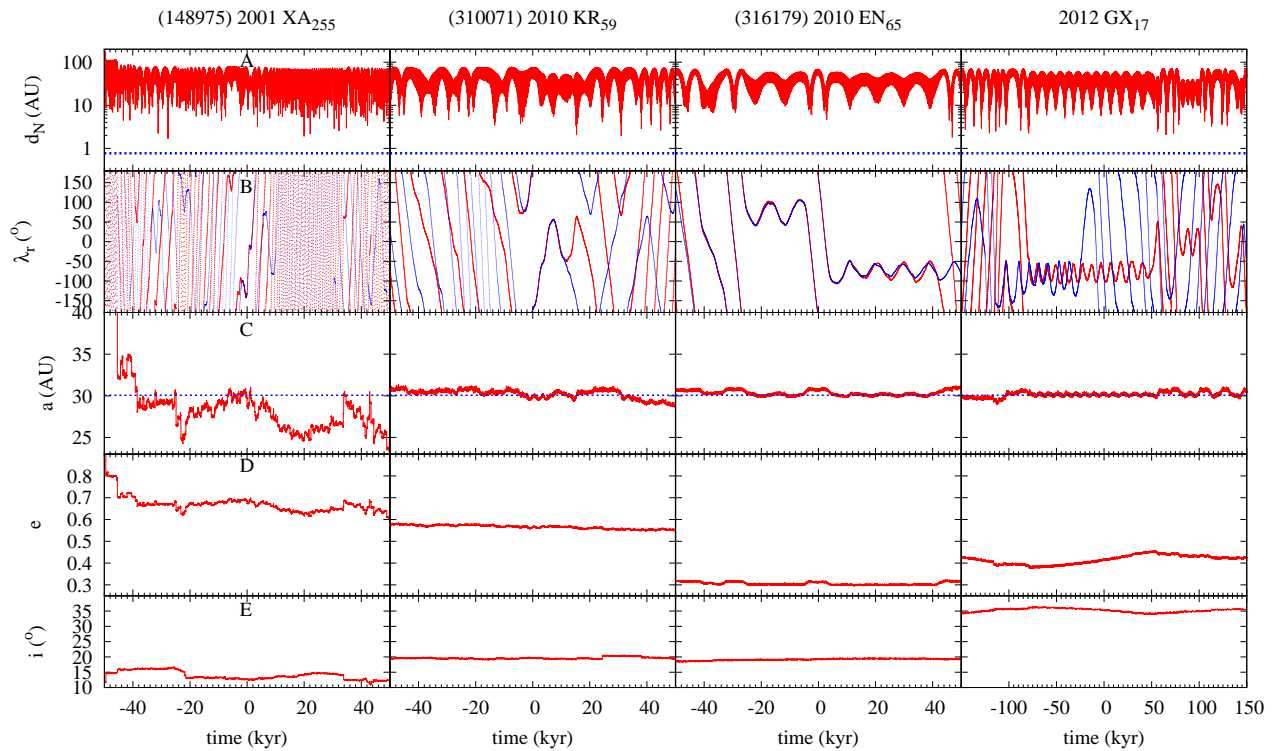}
        \caption{Time evolution of various parameters for 2001 XA$_{255}$, 2010 KR$_{59}$, 2010 EN$_{65}$, and 2012 GX$_{17}$. The 
                 distance to the object from Neptune, $d_N$, is shown in panel A and the value of the Hill sphere radius of 
                 Neptune, 0.769 AU, is displayed. The resonant angle, $\lambda_{r}$, is shown in panel B for the nominal orbit in 
                 Table \ref{elements} (thick red line) along with one of the control orbits (thin blue line). This particular 
                 control orbit was chosen to lie close to the 3-$\sigma$ limit so that its orbital elements are distinct from the 
                 nominal ones with the exception of 2012 GX$_{17}$. For that object the control orbit elements deviate by 1\% from 
                 the nominal values in Table \ref{elements}, which is at least 25 times higher than those of the other objects. 
                 The orbital elements are shown in panel C, $a$, with the current value of Neptune's semi-major axis, panel D, $e$, 
                 and panel E, $i$.
                }
        \label{all}
     \end{figure*}

      (148975) 2001 XA$_{255}$ was discovered on December 9, 2001 at Mauna Kea Observatory by Jewitt et al. (2002). It moves in a 
      very eccentric orbit (0.68) with perihelion just inside the orbit of Saturn and significant inclination (12.6$^{\circ}$). 
      Its aphelion is in the trans-Neptunian belt. It is classified as an inactive Centaur (Jewitt 2009). Its colours are neutral 
      (Fraser \& Brown 2012). (148975) 2001 XA$_{255}$ is the most dynamically unstable of the four objects studied here with an 
      e-folding time (or characteristic timescale on which two arbitrarily close orbits diverge exponentially) of about 300 yr. It 
      is perturbed by Saturn, Uranus, and Neptune. It appears to be a relatively recent (50 kyr) visitor from the scattered disk 
      on its way to the inner solar system. Calculations suggest that it came from beyond 100-200 AU and it is currently a very 
      short-lived horseshoe librator (see Figs. \ref{4orbits} and \ref{4alpha}). It entered the horseshoe orbital path about 10 
      kyr ago and will leave its present dynamical state 2 kyr from now (see Fig. \ref{all}); very close encounters with Neptune 
      are possible. It has previously been identified by Gallardo (2006) as a transition object affected by a 1:1 mean motion 
      resonance with Neptune and 1:2 with Uranus and was classified by Bailey \& Malhotra (2009) as a difussing Centaur. 
%
%
%
%

   \section{(310071) 2010 KR$_{59}$: a transient horseshoe}
      Discovered on May 18, 2010 by NASA's Wide-field Infrared Survey Explorer (WISE) (Scotti et al. 2010). The WISE telescope 
      scanned the entire sky in infrared light from January 2010 to February 2011 (Wright et al. 2010). (310071) 2010 KR$_{59}$ 
      moves in a very eccentric orbit (0.57) with perihelion outside Saturn's orbit and significant inclination (19.67$^{\circ}$). 
      Little is known about this object with the exception of its absolute magnitude of 7.7 ($\sim$22 apparent). This suggests a 
      relatively large object with an estimated diameter of about 100 km. Its e-folding time is close to 10 kyr, and it is now 
      pursuing a complicated horseshoe orbit (Fig. \ref{4alpha}) in its way to become a short-term quasi-satellite as depicted in 
      Fig. \ref{4orbits}. It will leave the 1:1 resonance with Neptune in a few 100 kyr. Its current horseshoe state started about 
      100 kyr ago. Dangerously close encounters with Neptune are possible.

   \section{(316179) 2010 EN$_{65}$: a jumping Trojan}
      (316179) 2010 EN$_{65}$ was discovered on March 7, 2010 by Rabinowitz and Tourtellotte using the 1.3-m reflector from Cerro 
      Tololo (Lowe et al. 2010). Upon discovery, the orbit of the object was described as Neptune-like, but its relative mean 
      longitude with respect to Neptune was almost 180$^{\circ}$. Up to 19 precovery images were found shortly after, the first 
      one acquired on November 4, 1989 by the Digitized Sky Survey (DSS) from Palomar Mountain. It is the largest of the four 
      objects studied here, perhaps with a diameter of nearly 200 km. It moves in an eccentric orbit (0.31) with perihelion just 
      outside Uranus's orbit and significant inclination (19$^{\circ}$). It was classified by Rabinowitz et al. (2012) as a 
      scattered-disk object. Our calculations show that its e-folding time is close to 3 kyr and it is currently moving from the 
      Lagrangian point L$_4$ to L$_5$ (see Figs. \ref{4orbits}, \ref{4alpha} and \ref{all}). This result is very robust as all 
      the control orbits follow the same orbital behaviour within 40 kyr of the current epoch. We classify this object as a 
      jumping Trojan.

   \section{2012 GX$_{17}$: a promising L$_5$ Trojan candidate}
      The minor planet 2012 GX$_{17}$ was discovered by Pan-STARRS 1, Haleakala, Hawaii on March 14, 2012 and was reobserved from 
      Magdalena Ridge Observatory, Socorro, New Mexico on March 18. As a recently discovered object and in sharp contrast with the
      previous three, its orbit is poorly known and it is mainly included here to encourage follow-up observations: its orbital
      solution is based on ten observations with an arc length of just four days. The physical properties of this object are 
      unknown with the exception of its absolute magnitude of 7.8 ($\sim$22 apparent). This suggests a relatively large object 
      with an estimated diameter in the range 55-180 km (for an albedo range of 0.5-0.05 and an assumed density of 2000 kg m$^{-3}$). 
      Its period, 165 yr, matches that of Neptune, 164.79 yr, therefore, it appears to follow a 1:1 resonant orbit with Neptune ($a$ = 
      30.13 AU), yet it has been classified as a Centaur by the MPC. Our calculations for the nominal orbit in Table \ref{elements} 
      indicate that the value of the relative mean longitude of 2012 GX$_{17}$ librates around -60$^{\circ}$ with an amplitude of 
      40-55$^{\circ}$ and a period of about 10 kyr; therefore, it appears to be a trailing Trojan (see Figs. \ref{4orbits}, 
      \ref{4alpha} and \ref{all}). Although its e-folding time is nearly 10 kyr, it has been an L$_5$ Trojan for tens of kyr and 
      simulations suggest that it will be leaving its current dynamical state to become a horseshoe librator in about 50 kyr. 2011 
      HM$_{102}$ was the highest inclination Neptune Trojan known but this candidate has even higher orbital inclination ($i$ = 
      35$^{\circ}$). Due to its poorly known orbit, we must insist that the object is a mere Trojan candidate (and in dire need of 
      follow-up observations) although all studied control orbits (with errors below 1\%) give consistent results. Our 
      calculations show that 2012 GX$_{17}$ may have been a passing minor body (perhaps from the scattered-disk) that eventually 
      evolved dynamically into a trailing Neptune Trojan. We also find that compared to Neptune's other Trojans, 2012 GX$_{17}$ is 
      significantly more perturbed by Uranus.  

   \section{Discussion and conclusions}
\begin{figure*}
       \centering
        \includegraphics[width=\textwidth]{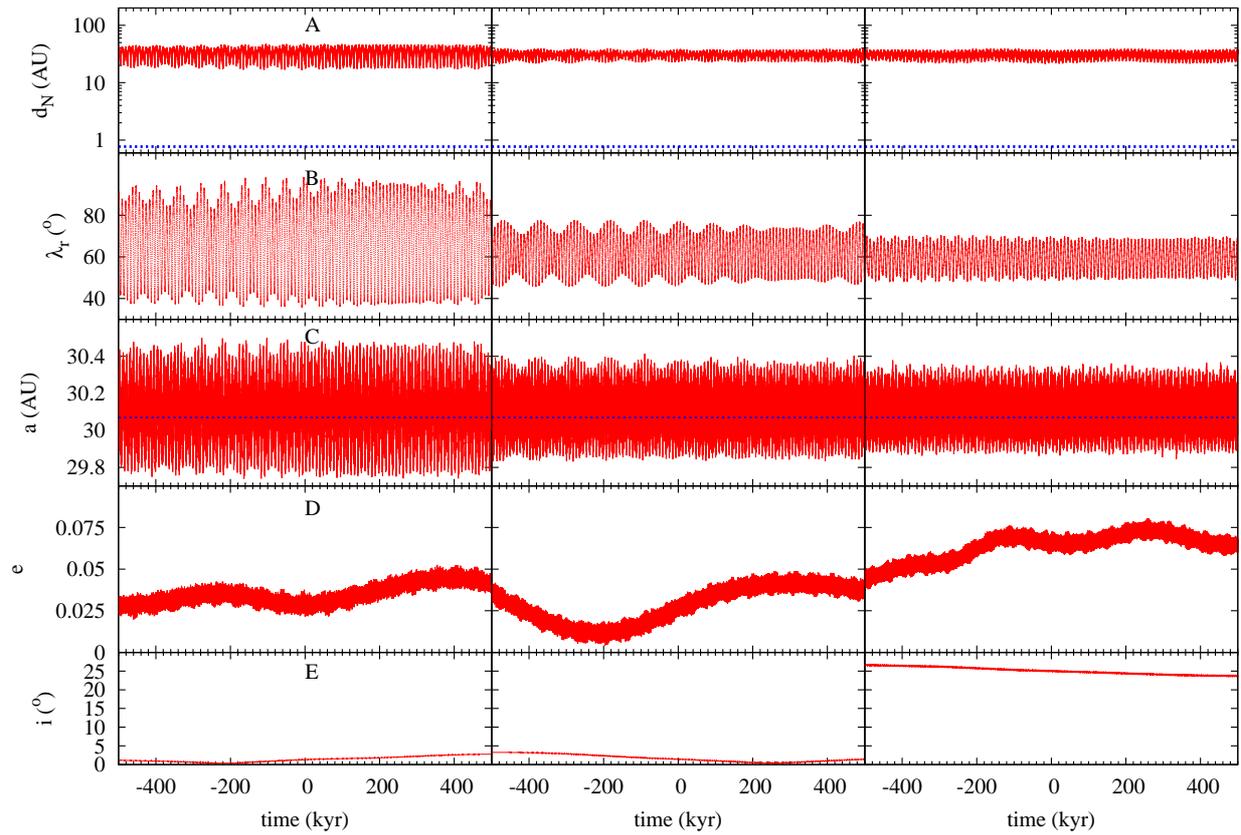}
        \caption{Evolution of various parameters during the time interval (-500, 500) kyr for three Neptune L$_4$ Trojans: 2001 
                 QR$_{322}$, 2004 UP$_{10}$, and 2005 TN$_{53}$. The distance to the objects from Neptune is given in panel A; the 
                 value of the Hill sphere radius of Neptune, 0.769 AU, is displayed. The resonant angle, $\lambda_{r}$, is 
                 displayed in panel B for the nominal orbit (source: JPL Small-Body Database and AstDyS-2). The orbital elements 
                 are depicted in panel C, $a$ with the current value of Neptune's semi-major axis, panel D, $e$, and panel E, $i$. 
                 2005 TN$_{53}$ appears to be the most stable of Neptune's Trojans and 2001 QR$_{322}$ the one with the largest 
                 libration amplitude.
                }
        \label{L4a}
     \end{figure*}

\begin{figure*}
       \centering
        \includegraphics[width=\textwidth]{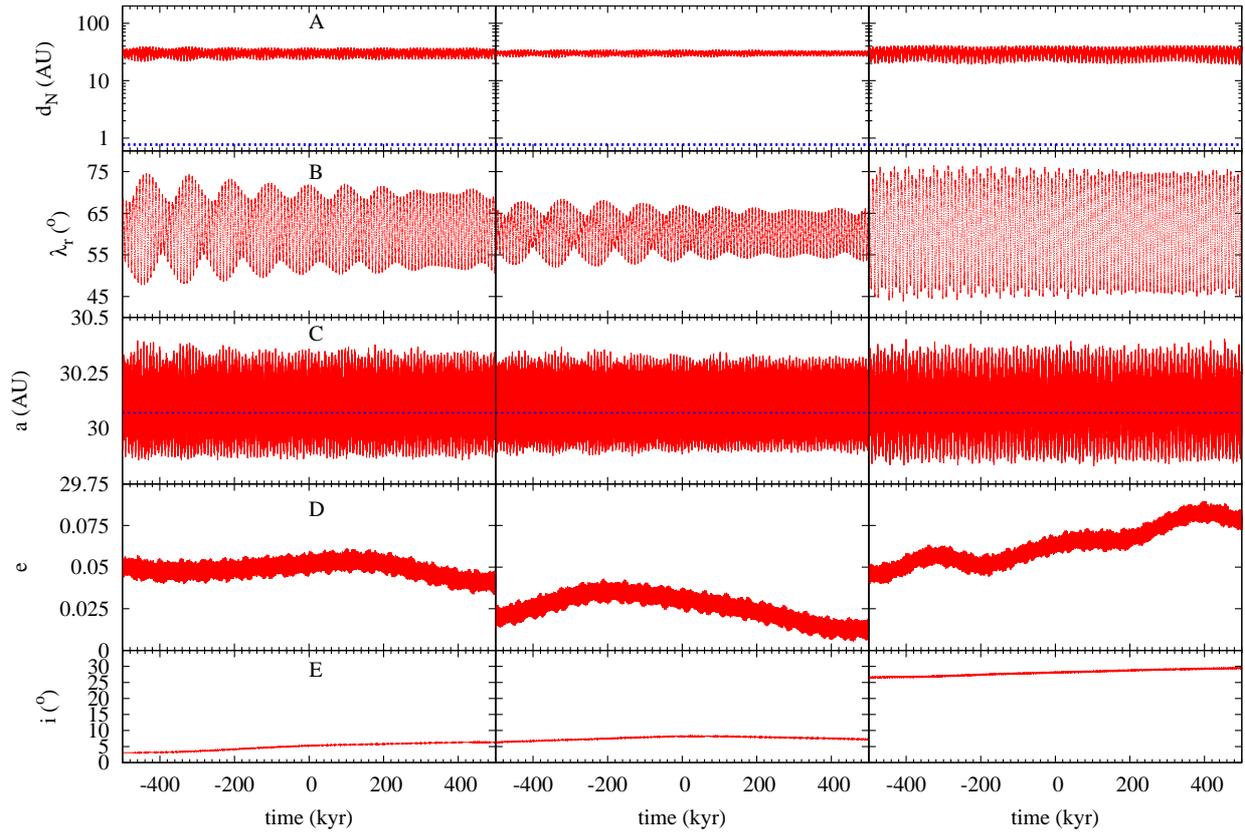}
        \caption{Same as Fig. \ref{L4a} but for Neptune L$_4$ Trojans: 2005 TO$_{74}$, 2006 RJ$_{103}$, and 2007 VL$_{305}$.
                 2007 VL$_{305}$ exhibits the second largest libration amplitude among L$_4$ Trojans.
                }
        \label{L4b}
     \end{figure*}

\begin{figure*}
       \centering
        \includegraphics[width=\textwidth]{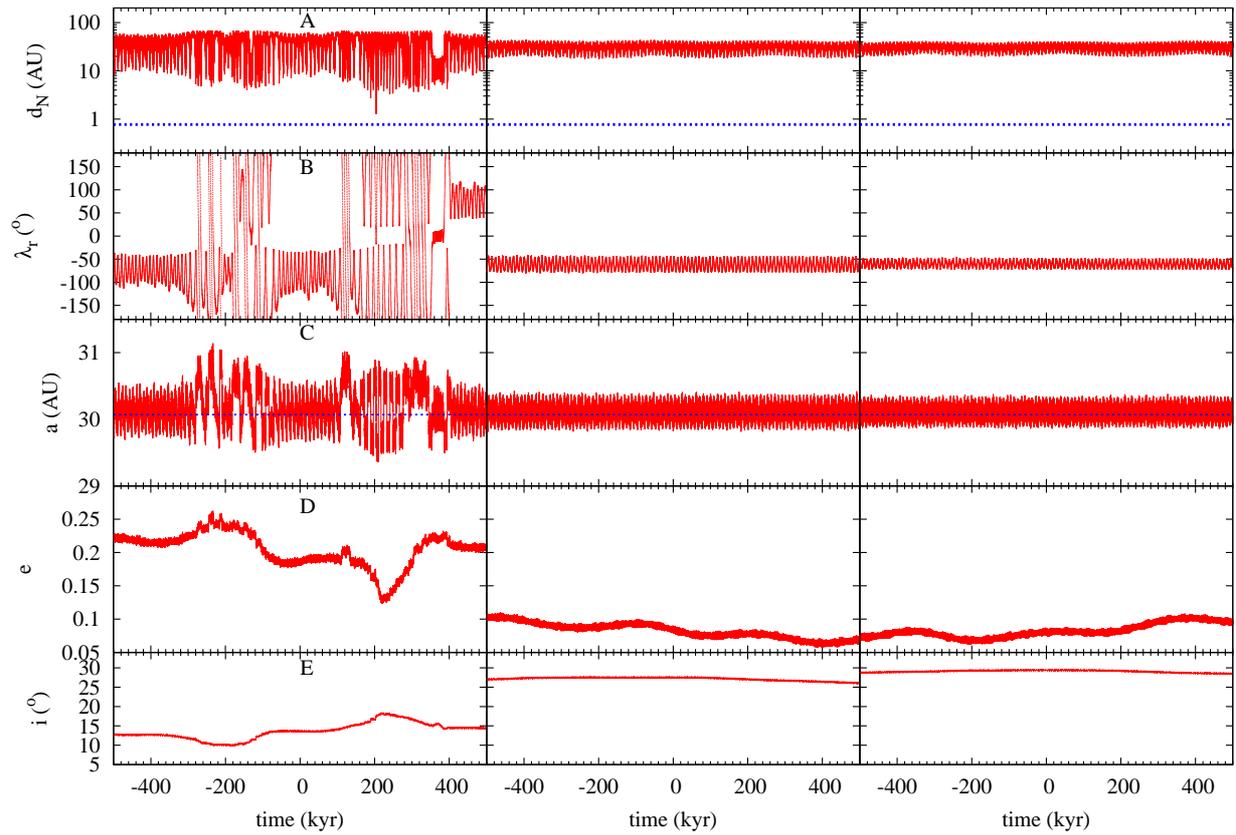}
        \caption{Same as Figs. \ref{L4a} and \ref{L4b} but for Neptune L$_5$ Trojans: 2004 KV$_{18}$, 2008 LC$_{18}$, and 2011
                 HM$_{102}$. The dynamical evolution of 2004 KV$_{18}$ is clearly different and it is the most unstable of the
                 "classical" Neptune Trojans. The most recent discovery, 2011 HM$_{102}$, is the most stable L$_5$ Trojan and it 
                 may be as stable as 2005 TN$_{53}$, likely the most stable of Neptune's Trojans.
                }
        \label{L5}
     \end{figure*}

      Compared to the eight "classical" Neptune Trojans (2001 QR$_{322}$, 2004 UP$_{10}$, 2005 TN$_{53}$, 2005 TO$_{74}$, 2006 
      RJ$_{103}$, 2007 VL$_{305}$, 2008 LC$_{18}$, and 2004 KV$_{18}$) and the new one (2011 HM$_{102}$), the objects described 
      here (even if they are currently co-orbitals of Neptune) are much more dynamically hot and their characteristic lifetime as 
      co-orbitals is relatively short, but comparable to that of the most unstable of the other nine objects (2004 KV$_{18}$, see 
      Figs. \ref{L4a}-\ref{L5}). Dynamically speaking, they are similar to the recently identified temporary Neptunian 
      quasi-satellite (309239) 2007 RW$_{10}$ (de la Fuente Marcos \& de la Fuente Marcos 2012a). The clue to the true origin of 
      all these five objects (and, perhaps, 2004 KV$_{18}$) could be (148975) 2001 XA$_{255}$, which clearly came from beyond 
      100-200 AU. This conclusion is similar to that in Horner \& Lykawka (2012) for the case of the Neptune L$_5$ Trojan 2004 
      KV$_{18}$. On the other hand, (316179) 2010 EN$_{65}$ could be the brightest and 2012 GX$_{17}$ the highest inclination
      Neptune Trojan.

      Out of more than 5,000 confirmed Jupiter Trojans there is a well-documented asymmetry between the number of bodies in the 
      leading and trailing populations: the Sloan Digital Sky Survey (SDSS) puts the L$_4$:L$_5$ Trojan ratio at 1.6$\pm$0.1 
      (Szab\'o et al. 2007), the Subaru telescope survey gives nearly 1.8 (Nakamura \& Yoshida 2008; Yoshida \& Nakamura 2008) 
      and the latest results from the WISE/NEOWISE mission give 1.4$\pm$0.2 (Grav et al. 2011). Prior to this research the 
      L$_4$:L$_5$ ratio for Neptune was 6/3=2. Our calculations suggest that Neptune's L$_5$ point is less stable than L$_4$, 
      apparently due to the influence of Pluto. All trailing Trojans undergo close approaches within 0.3 AU of Pluto. Close 
      encounters with the dwarf planet can send trailing Trojans into complex paths away from the usual tadpole orbit, inducing 
      jumping Trojan events, temporary quasi-satellite episodes, or throwing the object on a horseshoe orbit. An asymmetry between 
      the Neptunian L$_4$ and the L$_5$ swarms was found in numerical integrations by Holman \& Wisdom (1993) but, at that time, 
      it was not understood why that asymmetry existed. 
      
      It is sometimes claimed that Neptune cannot currently efficiently trap objects in the 1:1 commensurability even for short 
      periods of time. Our results argue that, contrary to this view, Neptune is still actively capturing temporary co-orbitals. 
      This is consistent with recent findings by Kortenkamp \& Joseph (2011). These authors conclude that the nearly 2:1 mean 
      motion resonance between Uranus and Neptune facilitates capture of new Neptune co-orbitals. Most of the objects described 
      here are relatively large and a much more numerous population of smaller objects similar to them may exist. We also confirm 
      that dynamical classifications based on cuts in the pericentre and other elements are not very reliable in the case of 
      resonant objects.

   \begin{acknowledgements}
      The authors would like to thank S. Aarseth for providing the code used in this research and the referee, S. Kortenkamp, for 
      his constructive and quick report. This work was partially supported by the Spanish 'Comunidad de Madrid' under grant CAM 
      S2009/ESP-1496 (Din\'amica Estelar y Sistemas Planetarios). We thank M.~J. Fern\'andez-Figueroa, M. Rego Fern\'andez and the 
      Department of Astrophysics of Universidad Complutense de Madrid (UCM) for providing excellent computing facilities. Most of 
      the calculations and part of the data analysis were completed on the 'Servidor Central de C\'alculo' of the UCM and we thank 
      S. Cano Als\'ua for his help during that stage. In preparation of this letter, we made use of the NASA Astrophysics Data 
      System and the ASTRO-PH e-print server.
   \end{acknowledgements}

%
%
         \begin{table*}
          \fontsize{8}{10pt}\selectfont
          \tabcolsep 0.35truecm
          \caption{Heliocentric Keplerian orbital elements of Neptune's L$_4$ Trojans (I): 2001 QR$_{322}$, 2004 UP$_{10}$, and 
                   2005 TN$_{53}$. Values include the 1-$\sigma$ uncertainty.
                   (Epoch = JD2456200.5, 2012-Sep-30.0; J2000.0 ecliptic and equinox.
                   Sources: JPL Small-Body Database and AstDyS-2.)
                  }
          \begin{tabular}{ccccc}
           \hline
                                                                   &   & 2001 QR$_{322}$     & 2004 UP$_{10}$      & 2005 TN$_{53}$       \\
           \hline
            semi-major axis, $a$ (AU)                              & = & 30.356$\pm$0.007    & 30.28$\pm$0.02      & 30.262$\pm$0.012     \\
            eccentricity, $e$                                      & = & 0.02891$\pm$0.00010 & 0.03140$\pm$0.00010 & 0.069$\pm$0.002      \\
            inclination, $i$ ($^{\circ}$)                          & = & 1.3210$\pm$0.0006   & 1.43107$\pm$0.00005 & 24.978$\pm$0.003     \\
            longitude of the ascending node, $\Omega$ ($^{\circ}$) & = & 151.548$\pm$0.015   & 34.798$\pm$0.003    & 9.2809$\pm$0.0002    \\
            argument of perihelion, $\omega$ ($^{\circ}$)          & = & 167.0$\pm$0.4       & 4$\pm$3             & 84.7$\pm$0.4         \\
            mean anomaly, $M$ ($^{\circ}$)                         & = & 60.3$\pm$0.4        & 345$\pm$2           & 296.7$\pm$0.2        \\
            perihelion distance, $q$ (AU)                          & = & 29.478$\pm$0.005    & 29.329$\pm$0.014    & 28.176$\pm$0.010     \\
            absolute magnitude, $H$ (mag)                          & = & 7.7                 & 8.8                 & 9.1                  \\
           \hline
          \end{tabular}
          \label{L4T1}
         \end{table*}
%
%
%
%
         \begin{table*}
          \fontsize{8}{10pt}\selectfont
          \tabcolsep 0.35truecm
          \caption{Heliocentric Keplerian orbital elements of Neptune's L$_4$ Trojans (II): 2005 TO$_{74}$, 2006 RJ$_{103}$, and 
                   2007 VL$_{305}$. Values include the 1-$\sigma$ uncertainty.
                   (Epoch = JD2456200.5, 2012-Sep-30.0; J2000.0 ecliptic and equinox.
                   Sources: JPL Small-Body Database and AstDyS-2.)
                  }
          \begin{tabular}{ccccc}
           \hline
                                   &   & 2005 TO$_{74}$    & 2006 RJ$_{103}$    & 2007 VL$_{305}$      \\
           \hline
            $a$ (AU)               & = & 30.262$\pm$0.013  & 30.181$\pm$0.009   & 30.195$\pm$0.014     \\
            $e$                    & = & 0.0536$\pm$0.0009 & 0.0300$\pm$0.0010  & 0.0684$\pm$0.0003    \\
            $i$ ($^{\circ}$)       & = & 5.248$\pm$0.002   & 8.1615$\pm$0.0003  & 28.1301$\pm$0.0010   \\
            $\Omega$ ($^{\circ}$)  & = & 169.370$\pm$0.006 & 120.912$\pm$0.010  & 188.5826$\pm$0.0010  \\
            $\omega$ ($^{\circ}$)  & = & 300.5$\pm$0.4     & 17$\pm$1           & 217.6$\pm$0.3        \\
            $M$ ($^{\circ}$)       & = & 278.7$\pm$0.4     & 257$\pm$1          & 359.2$\pm$0.2        \\
            $q$ (AU)               & = & 28.639$\pm$0.011  & 29.276$\pm$0.008   & 28.130$\pm$0.012     \\
            $H$ (mag)              & = & 8.5               & 7.5                & 8.0                  \\
           \hline
          \end{tabular}
          \label{L4T2}
         \end{table*}
%
%
%
%
         \begin{table*}
          \fontsize{8}{10pt}\selectfont
          \tabcolsep 0.35truecm
          \caption{Heliocentric Keplerian orbital elements of Neptune's L$_5$ Trojans: 2004 KV$_{18}$, 2008 LC$_{18}$, and 2011
                   HM$_{102}$. Values include the 1-$\sigma$ uncertainty.
                   (Epoch = JD2456200.5, 2012-Sep-30.0; J2000.0 ecliptic and equinox. 
                   Sources: JPL Small-Body Database and AstDyS-2.)
                  }
          \begin{tabular}{ccccc}
           \hline
                                   &   & 2004 KV$_{18}$        & 2008 LC$_{18}$       & 2011 HM$_{102}$       \\
           \hline
            $a$ (AU)               & = & 30.108$\pm$0.011      & 29.90$\pm$0.03       & 30.05$\pm$0.05        \\
            $e$                    & = & 0.1847$\pm$0.0011     & 0.086$\pm$0.003      & 0.0785$\pm$0.0002     \\
            $i$ ($^{\circ}$)       & = & 13.6127$\pm$0.0015    & 27.592$\pm$0.006     & 29.4203$\pm$0.0004    \\
            $\Omega$ ($^{\circ}$)  & = & 235.6308$\pm$0.0005   & 88.5219$\pm$0.0012   & 100.98900$\pm$0.00006 \\
            $\omega$ ($^{\circ}$)  & = & 294.2$\pm$0.3         & 8$\pm$13             & 152.05$\pm$0.03       \\
            $M$ ($^{\circ}$)       & = & 61.19$\pm$0.13        & 173$\pm$15           & 22.48$\pm$0.03        \\
            $q$ (AU)               & = & 24.548$\pm$0.007      & 27.327$\pm$0.007     & 27.691$\pm$0.004      \\
            $H$ (mag)              & = & 8.9                   & 8.4                  & 8.1                   \\
           \hline
          \end{tabular}
          \label{L5T}
         \end{table*}
%
%
%
%
%
%
%
%
%
%
%
%
%
%

\end{document}